\documentclass{article}
\usepackage{epsfig} % for postscript figures

 \newcommand{\eq}[1]{Eq.~(\ref{eq.#1})}	% Ref. to equation
 \newcommand{\fig}[1]{Fig.~\ref{fig.#1}}
 \newcommand{\eqlabel}[1]{\label{eq.#1}}

\newcommand{\figdef}[3]{% label, body, caption
\begin{figure}[!htb]
 \centering\leavevmode#2%
 \caption{#3}
 \label{fig.#1}
\end{figure}                 }

\title{\vspace{-0.7in}A Framework for Structured Quantum Search}

\author{\vspace{0.1in}Tad Hogg\\ Xerox Palo Alto
Research Center \\ 3333 Coyote Hill Road \\ Palo Alto, CA 94304,
U.S.A. \\ hogg@parc.xerox.com}

\begin{document}

\maketitle

\vspace{-0.1in}

\begin{abstract}
  A quantum algorithm for general combinatorial search that uses the
  underlying structure of the search space to increase the
  probability of finding a solution is presented.  This algorithm
  shows how coherent quantum systems can be matched to the
  underlying structure of abstract search spaces, and is analytically
  simpler than previous structured search methods.  The algorithm is
  evaluated empirically with a variety of search problems, and shown
  to be particularly effective for searches with many constraints.
  Furthermore, the algorithm provides a simple framework for
  utilizing search heuristics. It also exhibits the same phase
  transition in search difficulty as found for sophisticated
  classical search methods, indicating it is effectively using the
  problem structure.
\end{abstract}

\section{Introduction}

Combinatorial search problems are among the most difficult
computational problems because the time required to solve them often
grows exponentially with the size of the problem~\cite{garey79}.  Many
such problems have a great deal of structure, allowing heuristic
methods to greatly reduce the rate of exponential growth.  Quantum
computers~\cite{benioff82,bernstein92,deutsch85,deutsch89,divincenzo95,feynman86,lloyd93}
offer a new possibility for utilizing this structure with {\em quantum
parallelism}, i.e., the ability to operate simultaneously on a
superposition of many classical states, and {\em interference} among
different computational paths.

The physical restriction to unitary linear operations makes quantum
computers difficult program effectively.  Nevertheless, some
algorithms have been developed.  These include a method for
efficiently factoring integers~\cite{shor94}, a problem that appears
to be intractable for classical computers.  This method relies on
rapidly identifying periods of periodic functions, so is limited to
problems that can be cast in this form.  More recently, a general
search method was proposed~\cite{grover96}.  While a substantial
improvement over classical search for unstructured search spaces, it
ignores the structure found in many combinatorial search problems thus
limiting its effectiveness for such cases.  In many such problems,
solutions can be built incrementally from smaller parts, resulting in
substantial improvement in classical searches through the use of
heuristics that exploit this property.  This observation forms the
basis for a quantum search algorithm that uses structure in much the
same way as classical heuristic searches~\cite{hogg95d}.

These general search algorithms operate with superpositions of all
possible search states for the problem.  Each of their steps consists
of adjusting the phases of the amplitudes in the superposition, based
on properties of the problem being solved, combined with a
problem-independent operation to mix amplitudes among different
states. These algorithms are probabilistic and incomplete: they are
likely to find solutions if any exist, but cannot guarantee no
solutions exist. As with classical searches, the number of consistency
tests (or checks) required by the algorithm characterizes search cost,
although the detailed cost of each step will vary somewhat among
different algorithms and implementations.

Specifically, the unstructured algorithm is likely to find one of $S$
solutions among $N$ possibilities with $O(\sqrt{N/S})$
checks~\cite{boyer96,grover96}. Without using additional structure,
the fastest classical search is generate-and-test, where possible
solutions are examined sequentially and which requires $O(N/S)$
checks. Thus this quantum algorithm is a substantial improvement, and
is close to the best possible for unstructured quantum
searches~\cite{boyer96}. 

The structured algorithm~\cite{hogg95d} builds solutions
incrementally. Each trial requires only $O(\log N)$ checks, but gives
a low probability to find a solution, thus requiring multiple
trials. The number of trials required is difficult to determine
theoretically since the use of problem structure introduces many
dependencies among the steps. Instead, as with many classical
heuristic searches, the algorithm was evaluated empirically, which is
necessarily limited to small problems due to the exponential slowdown
associated with classical simulations of quantum computations. These
simulations demonstrate a substantial improvement, on average, for
classes of random problems. 

While these algorithms are encouraging developments, the extent to
which quantum searches can improve on heuristically guided classical
methods for structured problems remains an open question.  Even if
quantum computers are not applicable to {\em all} combinatorial search
problems, they may still be useful for many instances encountered in
practice. This is an important distinction since typical instances of
search problems are often much easier to solve than is suggested by
worst case analyses, though even typical costs often grow
exponentially on classical machines.  The study of the average or
typical behavior of search heuristics relies primarily on empirical
evaluation. This is because the complicated conditional dependencies
in search choices made by the heuristic often preclude a simple
theoretical analysis, although phenomenological theories can give an
approximate description of some generic
behaviors~\cite{hogg94b,kirkpatrick94,monasson96}.

In fact, the hard instances are not only rare but also concentrated
near abrupt transitions in problem behavior analogous to physical
phase transitions~\cite{cheeseman92,hogg94b,web.hogg94}.  These
transitions correspond to a change from underconstrained to
overconstrained problems and reflect changes in the structure of the
problems.  Problems located away from the transition region, i.e.,
with relatively few or relatively many constraints, tend to be easy to
solve. These transitions appear with many classical search methods
that use problem structure to guide choices, independent of the
detailed nature of the search procedure. They thus reflect a universal
property of classes of search problems rather than specific search
algorithms.Similarly, the structured quantum algorithm also exhibits
this transition.  By contrast, the performance of unstructured search
methods, such as generate-and-test and the unstructured quantum search
algorithm, varies only with the number of solutions a problem has.
Thus unstructured methods do not exhibit the transition, and in
particular their search cost does not decrease when applied to
increasingly constrained search problems.  Thus an indication of
whether a quantum algorithm exploits problem structure, through
interference among different computational paths, is whether it
exhibits the transition behavior.

In this paper, a new and analytically simpler structured quantum
search algorithm is presented.  Specifically the following two
sections review the underlying structure of many combinatorial search
problems and how structure can be used with quantum superpositions.
The new algorithm is then described followed by an evaluation of its
behavior. Finally some open issues are described. The structure-based
algorithm provides a framework within which additional heuristics with
knowledge of the structure of specific problems can be
incorporated. It thus provides a way to develop and evaluate the use
of heuristics for quantum searches, in a manner analogous to the use
of heuristics to dramatically improve many classical search
strategies.

\section{The Structure of Combinatorial Search}

NP search problems have exponentially many possible states and a
procedure that quickly checks whether a given state is a
solution~\cite{garey79}.  Constraint satisfaction problems
(CSPs)~\cite{mackworth92} are an important example.  A CSP consists of
$\nu$ variables, $V_1,\ldots,V_\nu$, and the requirement to assign a
value to each variable to satisfy given constraints. Searches examine
various {\em assignments}, which give values to some of the
variables. {\em Complete} assignments have a value for every variable.
Search states can also be viewed as sets of {\em assumptions}, where
an assumption is an assignment to a single variable, e.g., $V_1=0$.

More generally, combinatorial search can be viewed as finding, from
among $n$ given assumptions, a set of size $L$ satisfying specified
constraints.  Such a set is a solution to the problem.  Sets of
assumptions that violate a constraint are {\em nogoods}. In the
particular case of CSPs, these include the {\em necessary nogoods}, in
which some variables are assigned multiple
values~\cite{williams92}. The remaining sets are goods, i.e.,
consistent. Supersets of a nogood are also nogood so sets of
assumptions are usefully viewed as forming a lattice with levels from
0 to $n$, with level $i$ containing all sets of size $i$. This
lattice, describing the consistency relationships among sets, is the
{\em deep structure} of the combinatorial search problem.  This
structure for $n=4$ is shown in \fig{lattice}.  Notationally, we
denote the size of a set $s$ by $|s|$.

\figdef{lattice}{
\epsfig{file=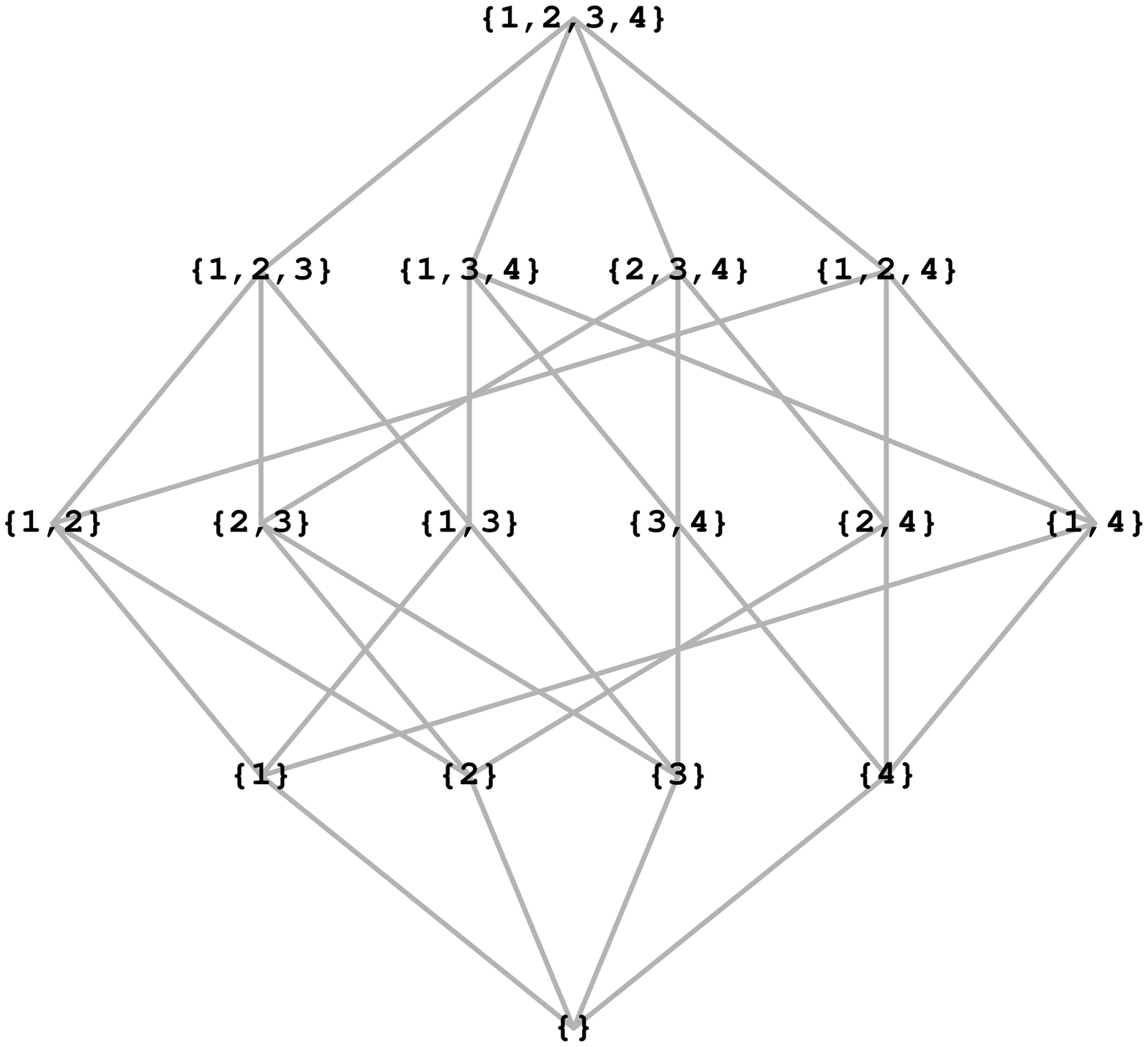,width=3.5in}
}{Set lattice for a problem with four assumptions, containing all subsets
 of $\{ 1,2,3,4 \}$. The bottom of the lattice, level 0, represents the 
single set of size zero, the four points at level 1 represent the four singleton subsets, etc.}

Classically, the necessary nogoods for CSPs can be avoided completely by
searching only among assignments. Unfortunately, no quantum procedure
can incrementally produce complete assignments from smaller ones with
the variety of variable orderings needed for effective
search~\cite{hogg95d}. Thus incremental quantum algorithms must use
the expanded search space containing necessary nogoods.

This abstract description of combinatorial search in terms of sets of
assumptions is less commonly used than other representations, which
are more compact and efficient for classical search algorithms. It is
introduced here as a useful basis for quantum searches and because it
applies to many search problems, including CSPs. 

Important examples of CSPs are graph coloring and satisfiability.
In coloring an $\nu$-node graph with $c$ colors an assumption
$V=\kappa$ is an assignment of a color $\kappa$ to a node $V$.  Thus
there are $n = \nu c$ assumptions for this problem and a solution is a
set of $L=\nu$ such assumptions that gives a unique color to each node
(i.e., contains no necessary nogood as a subset) and distinct colors
to each pair of nodes in the graph that are linked by an edge. Each
edge is a constraint directly specifying $c$ nogoods, each consisting
of a pair of assumptions with the same color for both of the nodes
linked by that edge.  This search problem is known to be NP-complete
for a fixed $c$ (at least equal to 3) as $\nu$ grows.  

The satisfiability problem consists of a propositional formula in
$\nu$ variables and the requirement to find a value (true or false)
for each variable that makes the formula true. This problem has
$n=2\nu$ assumptions and $L=\nu$. An NP-complete example is 3-SAT,
where the formula consists of a conjunction of clauses, and each
clause is a disjunction of 3 of the variables in the problem, any of
which may be negated. Thus a solution must satisfy every clause in the
formula. An example of such a clause, with the third variable negated,
is $V_1$ OR $V_2$ OR $\overline{V_3}$, which is false for exactly one
assignment for these variables: $\{V_1={\rm false}, V_2={\rm false},
V_3={\rm true}\}$. Thus each clause introduces a single nogood of size
3.

These examples show that challenging examples of combinatorial search
occur when the nogoods directly determined by the constraints have a
fixed size while the number of assumptions and the size of solutions
grows linearly with problem size. This scaling, which gives a high
concentration of hard instances~\cite{williams92}, is used in the
experiments described below.

\section{Using Structure for Quantum Search}

How can problem structure be used to improve quantum search?  Some
suggestions are provided by three categories of classical methods that
use different aspects of problem structure. These methods contrast
with unstructured searches, which amount to a random search among the
states or a systematic enumeration of them without any use of prior
results to guide future choices.

First, the problem can be simplified or {\em abstracted} in some way.
A solution to the abstract problem is then used to guide the search in
the original problem. If the abstract problem can be solved rapidly
and its solution provides a good starting point for the full
search, this strategy can be effective.  For CSPs, abstraction can
consist of ignoring some of the constraints or identifying useful
hierarchical aggregations of variables and constraints.

A second method takes advantage of the clustering of solutions found
in many search problems. That is, instead of being randomly
distributed throughout the search space, the states have a simple
neighborhood relationship such that states with a few or no conflicts
tend to be near other such states. This neighborhood relationship is
used by {\em repair} searches. Starting from a random state, they
repeatedly select from among the current state's neighbors one that
reduces the number of conflicts with the constraints. Such searches
can become stuck in local minima but are often very
effective~\cite{minton92,selman92}. More sophisticated versions
address the problem of local minima by allowing occasional changes
that increase the number of conflicts~\cite{kirkpatrick83} as well as
using a population of search states and combining parts from those
with relatively few conflicts~\cite{davis87}.  For example, with CSPs,
the neighbors of a given complete assignment could be all other
assignments with a different value for just one variable.

The third general search category {\em builds} solutions incrementally
from smaller parts, which requires expanding the overall
search space to include these smaller parts. These
methods exploit the fact that in many problems the small
parts can be tested for consistency before they are expanded to
complete states. When a small state is found to be inconsistent,
all possible extensions of it will also be inconsistent,
allowing an early pruning of the search. In such cases, the search
backtracks to a prior decision point to try a different
incremental construction. For CSPs, this method assigns values to
variables one at a time until a conflict is found. Its performance can
be very good, but depends greatly on the choice of the order in which
the variables are considered: a poor choice can mean few opportunities
for early pruning.

In the context of quantum search, the general aspects of these methods
could be used by a unitary mapping that, at least approximately, maps
amplitude from one state to others that would be considered after it
by the corresponding classical method. In effect, this allows
examining, and using interference from, all possible choices the
classical search could have made, rather than being restricted to a
single series of choices at a time. The details of the particular
problem being solved could be introduced by adjustments to the phases
of the amplitudes based on testing states for consistency. This
technique, used with both the unstructured and structured quantum
searches mentioned above, neatly separates the design of the unitary
matrix that mixes amplitudes from any consideration of the detailed
nature of specific problems. In particular, the structured algorithm
builds solutions incrementally, but is difficult to analyze
theoretically because the matrix elements used in the mapping
procedure must be evaluated numerically. Thus it is of interest to see
if there are analytically simpler structured methods that nevertheless
retain the same properties of concentrating amplitude into solutions
incrementally.

Toward this end we can consider the necessary size of the matrix
elements connecting different sets. Consider the diffusion matrix used
in the unstructured search algorithm~\cite{grover96}. Its off-diagonal
terms have magnitude of size $O(1/N)$. If there is a single solution
and we start from a uniform initial state with amplitudes
$1/\sqrt{N}$, then even with a perfect choice of phases so each set
gives a positive contribution to the solution, each step adds $O((1/N)
N/\sqrt{N}) = O(1/\sqrt{N})$ to the solution set, because there are
$N-1$ nonsolutions each mapped to the solution by a matrix element of
size $O(1/N)$. Thus $O(\sqrt{N})$ steps will be required to give a
solution amplitude of $O(1)$. This informal argument corresponds to
that from a more detailed analysis of the unstructured
algorithm~\cite{grover96,boyer96}. Starting with other initial
conditions does not improve the situation, e.g., if all amplitude is
initially in a single set, each step contributes $O(1/N)$ to the
solution set, requiring $O(N)$ steps.  Thus, any substantial
improvement in search cost requires matrix elements with much larger
couplings between sets. In particular, suppose each set receives
substantial contributions from $z$ other sets with matrix elements of
size $u$. Then normalization requires that $z u^2 \leq O(1)$. A single
step, again assuming the phases are chosen perfectly, can transfer
$O(u z/\sqrt{z}) = O(1)$ to a single set starting from a uniform
distribution among the $z$ sets, or $O(u) = O(1/\sqrt{z})$ starting
from all amplitude in a single set.  This observation suggests that at
least some matrix elements must be of size equal to a power of
$O(1/\log N)$ to have a good chance of moving significant amplitude
from one group of sets to a solution in a power of $O(\log N)$
steps. Such a matrix would be used to rapidly transfer amplitude from
one group of sets to another, where each group is more likely to
include solutions. Whether this performance can be realized depends on
how well the phases can actually be chosen to give positive
contribution to solutions and how the different groups of sets are
selected. In the previous structured algorithm~\cite{hogg95d}, and the
new one presented below, these choices are based on searches that
incrementally construct solutions from smaller parts.

\section{A Structure-Based Quantum Search}

A search with $n$ assumptions operates in the full lattice with
$N=2^n$ sets.  Let $\psi^{( j)}_{s}$ be the amplitude of the set $s$
after completing step $j$ of the algorithm.  A single trial of the
search algorithm consists of:
\begin{enumerate}
\item initialize all amplitude in the empty set, i.e.,
  $\psi^{(0)}_s=1$ if the set $s=\emptyset$, and otherwise is zero.

\item iterate: for step $j$ from 1 to $J$, adjust phases based on
  consistency and then multiply by the matrix $U$ described below,
  to give
\begin{equation}\eqlabel{map}
\psi^{(j)}_r = \sum_s U_{rs} \rho_s \psi^{(j-1)}_s
\end{equation}
where $\rho_s$ is the phase assigned to the set $s$ as described below.

\item measure the final superposition
\end{enumerate}

The phase adjustment is the only part of the algorithm that depends
on the particular problem instance being solved.  The
choice of phase should move amplitude from nogoods to goods and also
toward the largest goods, which are the solutions to the problem. In
the previous structured and unstructured
algorithms~\cite{grover96,hogg95d}, this phase choice consisted simply
in inverting the phase of all nogoods, i.e., using $\rho _{s}=1$ when
$s$ is a good, and otherwise $\rho _{s}=-1$. This is an effective
choice for the algorithm described here.  However, performance is
somewhat better in all but very highly constrained problems if, at
each step, successively larger goods also have their phases inverted, i.e.,
\begin{equation}\eqlabel{phase}
\rho_s = \left\{ \begin{array}{rl}
	-1	& \mbox{if $s$ is nogood or $|s|<\min(L,j-1)$} \\
	1	& \mbox{otherwise}
		\end{array}
	\right.
\end{equation}

After $J$ steps, the final measurement gives a single set. This set
will be a solution with probability
\begin{equation}\eqlabel{p(soln)}
P_{\rm soln}=\sum _{s}p( s)
\end{equation}
with the sum over solution sets. Here $p( s)={\left|
\psi^{(J)}_{s}\right| }^{2}$ is the probability to obtain the set $s$
with the measurement of the final state. On average, the algorithm
will need to be repeated $T=1/P_{\rm soln}$ times to find a solution.

The search cost can be characterized by the number of steps
required to find a solution on average, i.e., $C = JT$. As
described below, the matrix $U$ emphasizes mapping amplitude from sets
to their supersets with one additional assumption. So one might expect
the algorithm would require $L$ steps to give significant amplitude to
sets of size $L$. However, the experiments reported below show fairly
large amplitudes with somewhat fewer steps. In addition, instead of
continuing the algorithm to maximize the probability to have a
solution, a lower average search cost is sometimes possible by
stopping earlier~\cite{boyer96}, a simple strategy for improving
probabilistic algorithms~\cite{luby93}. Determining the best number of
steps to take remains an open problem, but at worst one could try the
algorithm for all values of $J$ up to $L$, resulting in at worst a
linear increase in the overall search cost because
$L \leq n$. More sophisticated methods for finding a suitable number of
steps to take have been proposed for the unstructured search
algorithm~\cite{boyer96} and similar techniques may be useful for this
structured search as well.

\subsection{A Structure-Based Mapping}

The matrix $U$, mixing amplitudes from different states, is the part
of the algorithm that exploits structure to focus amplitude toward
solutions. Specifically, let $U = W D W$ where, for sets $r$ and $s$,
\begin{equation}
W_{rs} = \frac{1}{\sqrt{N}} (-1)^{|r \cap s|}
\end{equation}
and $D$ is a diagonal matrix of phases (complex numbers with magnitude
equal to 1) depending only on the size of the sets, i.e., $D_{rr} =
d_{|r|}$. The matrix $U$ is readily shown to be unitary and its
multiplication of state vectors can be done rapidly on quantum
computers using a recursive decomposition of the matrix
$W$~\cite{boyer96,grover96}.

To take advantage of the lattice structure to incrementally construct
solutions, the elements of $U$ mapping from a set to its supersets
with one more item should be as large as possible. This can be done through
appropriate choice of the values of $d_k$. Specifically,
\begin{equation}
U_{rs} = \frac{1}{N} \sum_{k=0}^n d_k S_k(r,s)
\end{equation}
where
\begin{equation}
S_k(r,s) = \sum_{t, |t|=k} (-1)^{|r \cap t| + |s \cap t|}
\end{equation}
with the sum over all sets $t$ of size $k$.

A given element $e$ of $t$ contributes 0, 1 or 2 to $|r \cap t| + |s
\cap t|$ when $e$ is in neither $r$ nor $s$, in exactly one of $r$ or
$s$, or in both $r$ and $s$, respectively. Thus $(-1)^{|r \cap t| + |s
\cap t|}$ equals $(-1)^\lambda$ where $\lambda$ is the number of
elements in $t$ that are in exactly one of $r$ and $s$. There are
$(|r|-|r \cap s|) + (|s|-|r \cap s|)$ assumptions from which such elements
of $t$ can be selected.  Thus the number of sets $t$ of size $k$ with
$\lambda$ elements in exactly one of $r$ and $s$ is given by ${m
\choose \lambda} {n-m \choose k-\lambda}$ where $m = |r|+|s|-2|r \cap
s|$. Thus $S_k(r,s) = S_{km}^{(n)}$ where
\begin{equation}\eqlabel{Skm}
S_{km}^{(n)} = \sum_{\lambda} (-1)^\lambda {m \choose \lambda} {n-m \choose k-\lambda}
\end{equation}
so that $U_{rs}=\sum_k d_k S_{km}^{(n)}/N \equiv u_m$.

When $s$ is an immediate subset of $r$, i.e., $s \subset r$ and
$|r|=|s|+1$, we have $m=1$.
%Conversely, if $m=1$ then $s$ is an immediate subset or $r$ or vice versa.
%To see this, consider first $|r|=|s|$.
%Then $m = 2(|s|-|r \cap s|)$ which is even so cannot equal 1.
%When $|r|>|s|$, $m > 2(|s|-|r \cap s|)$ so when $m=1$ this lower bound must be
%zero since it is even, i.e., we must have $|s|=|r \cap s|$ so $s \subset r$.
%In this case $m = |r|-|s|$ which requires $|r|=|s|+1$ when $m=1$.
%Similarly when $|r|<|s|$.
Thus the value of $u_1$ governs the mapping of amplitudes from sets to
their immediate supersets and subsets.  To select the values of $d_k$
that maximize $u_1$, note that
$S_{01}^{(n)} = 1$ and $S_{k1}^{(n)} = {n-1 \choose k} - {n-1 \choose k-1}$.
% Using the relation ${n-1 \choose k} = {n-1 \choose k-1} (n-k)/k$ for $k>0$
% gives $S_{k1}^{(n)} = {n-1 \choose k-1} (n-2k)/k$. 
Thus $S_{k1}^{(n)}$ is positive for $n>2k$ and negative for $n<2k$, and
$u_1$ is maximized by selecting $d_k$ to be 1 for $k<n/2$
and $-1$ for $k>n/2$. If $n$ is even, $S_{k1}^{(n)}$ is zero for $k=n/2$ so
the choice of $d_{n/2}$ does not affect the value of $u_1$, though it
does affect other matrix elements.  In this case, we take $d_{n/2}=1$.
These choices give $u_1 = \frac{2}{N} {n-1 \choose \lfloor n/2
\rfloor}$ which scales as $\sqrt{2/(\pi n)}$ as $n \rightarrow
\infty$. Note this is much larger than the off-diagonal matrix
elements in the diffusion matrix used in the unstructured search
algorithm~\cite{grover96}, which are $O(1/N) = O(2^{-n})$.  Unlike the
previous structured search~\cite{hogg95d}, $U$ also gives some mixing
among sets separated by more than one level in the lattice.

\subsection{Classical Simulation}

As a practical matter, it is helpful if a quantum search method can be
evaluated effectively on existing classical computers.
Unfortunately, the exponential slowdown and growth in memory required
for such a simulation severely limits the size of feasible 
problems. For example, \eq{map} is a matrix
multiplication of a vector of size $2^n$ so a direct evaluation
requires $O(2^{2n})$ multiplications.

For the algorithm presented here, the cost of the classical simulation 
can be reduced substantially by
exploiting the map{'}s simple structure.
Specifically, the product $W{\bf x}$ can be computed recursively. To
see this consider the sets ordered by the value of the integer with
corresponding binary representation, e.g., the sets without item $n$
come before those with $n$. For example, the sets for $n=3$ are
ordered as $\{\}$, $\{1\}$, $\{2\}$, $\{1,2\}$, $\{3\}$, $\{1,3\}$,
$\{2,3\}$ and $\{1,2,3\}$. In this ordering, the matrix $W$ has the
recursive decomposition
\begin{equation}
W = \left( 
	\matrix{
		W^{\prime}&W^{\prime}\cr
		W^{\prime}&-W^{\prime}\cr
	}
\right)
\end{equation}
where $W^{\prime}$ is the same matrix but defined on subsets of
$\{ 1,\ldots ,n-1\}$.
We can then compute
\begin{equation}
W{\bf x} = \left(
	\matrix{
		W^{\prime}{\bf x^{(1)}} + W^{\prime}{\bf x^{(2)}} \cr
		W^{\prime}{\bf x^{(1)}} - W^{\prime}{\bf x^{(2)}} \cr
	}
\right)
\end{equation}
where $\bf x^{(1)}$ and $\bf x^{(2)}$ denote, respectively, the
first and second halves of the vector $\bf x$ (i.e., corresponding to
sets without $n$ and with $n$ respectively). Thus the cost to compute
$W{\bf x}$ is $C(n)=2C(n-1)+O(2^n)$ resulting in an
overall cost of order $n2^n$. While still exponential, this improves
substantially on the cost for the direct evaluation on classical machines.

\section{Quantum Search Behavior}

The behavior of this search algorithm was examined through a classical
simulation. While these results are limited to small problems,
they nevertheless give an indication of how this algorithm can
dramatically focus amplitude into solutions. As a check on the
numerical errors, the norm of the state vector remained within
$10^{-10}$ of 1.

\subsection{Extreme Cases}

The simplest examples are the extreme cases of problems with the
minimum and maximum possible number of nogoods. These cases have a
very uniform consistency structure and may be particularly useful
for analytic treatment of the algorithm. However, these problems are also
rather easy for classical methods.

The {\em minimum nogood problem} consists of having all sets of size less
than or equal to $L$ be goods, and all larger sets nogoods. Thus every
set of size $L$ is a solution. Classical or quantum methods that
operate only with complete sets (i.e., sets of size $L$)
will find a solution in a single try. Classical incremental methods
will require $L$ steps to construct a solution. Since $L=O(n)$, either
type of search can solve this problem rapidly. For the structured
quantum method, the amplitude of a set will depend only on the size of
the set, i.e., $\psi_s = \psi_{|s|}$. Thus \eq{map} becomes, for $h$
and $k$ running from 0 to $n$,
\begin{equation}
\psi^{(j+1)}_h = \sum_k V^{\rm min}_{hk} \rho_k \psi^{(j)}_k
\end{equation}
and \eq{phase} becomes $\rho_k=-1$ when either $k>L$ (i.e., the
corresponding sets are nogood) or $k<\min(j-1,L)$. The matrix in this
mapping is $V^{\rm min} = W^{\rm min} D^{\rm min} W^{\rm min}$ where
$D^{\rm min}$ is a diagonal matrix with $D^{\rm min}_{kk}$ equal to 1
for $k \leq n/2$ and $-1$ otherwise, and
\begin{equation}
W^{\rm min}_{hk} = \frac{1}{\sqrt{N}}\sum_z (-1)^z {h \choose z} {n-h \choose k-z} = \frac{S_{kh}^{(n)}}{\sqrt{N}}
\end{equation}
from \eq{Skm}.  Here the binomials in the sum count, for a set $r$ of
size $h$, the number of sets $s$ of size $k$ that have $z$ elements in
common with $r$.

At the other extreme, the {\em maximum nogood problem} consists of having a
single set of size $L$ and its subsets as goods and all other sets in
the lattice as nogoods. This problem thus has a single solution, which
without loss of generality we can take to be the set
$\{1,\ldots,L\}$. In this case, search methods that operate with
complete states will require more steps to find the single solution
out of the total of $n \choose L$ complete states. However, the large
number of nogoods will often allow classical heuristic repair methods
to find a solution rapidly. Similarly, incremental classical methods
will encounter conflicts immediately upon adding any assumption that
is not a subset of the solution, thus allowing the solution to be
found in $O(L)$ steps. For the structured quantum method, the
amplitude of a set will depend only on the size of the set and its
overlap with the single solution, i.e., $\psi_s = \psi_{|s|,|s \cap
\{1,\ldots,L\}|}$. Hence the state can be represented by a doubly
indexed vector $\psi_{kl}$ where $k$, giving the size of the set,
ranges from 0 to $n$ and $l$, giving its overlap with the solution,
ranges from $\max(0,k-(n-L))$ to $\min(k,L)$.  Thus \eq{map} becomes
\begin{equation}
\psi^{(j+1)}_{h j}= \sum_{kl} V^{\rm max}_{hj,kl} \rho_{kl} \psi^{(j)}_{kl}
\end{equation}
and \eq{phase} gives $\rho_{kl}=-1$ when either $k>L$ or $l<k$ (i.e.,
the corresponding sets are nogood) or $k<\min(j-1,L)$. The matrix in
this mapping is $V^{\rm max} = W^{\rm max} D^{\rm max} W^{\rm max}$
where $D^{\rm max}$ is a diagonal matrix with $D^{\rm max}_{kl,kl}$
equal to 1 for $k \leq n/2$ and $-1$ otherwise, and
\begin{equation}
W^{\rm max}_{hj,kl} = \frac{1}{\sqrt{N}} \sum_{zx} (-1)^z {L-j \choose x} {j \choose l-x}
{h-j \choose z-l+x} {n-L-h+j \choose k-z-x}
\end{equation}
In this double sum, the binomials count, for a set $r$ of size $h$
with $j$ elements in common with the single solution, the number of
sets $s$ of size $k$, with $l$ elements in the single solution, $z$
elements in common with $r$ and $x$ elements in the solution
but not in the set $r$. This double sum separates to give
\begin{equation}
W^{\rm max}_{hj,kl} = \frac{1}{\sqrt{N}} S^{(L)}_{jl} S^{(n-L)}_{h-j,k-l}
\end{equation}
from \eq{Skm}.

The scaling of the search cost for these extreme problems is shown in
\fig{x:extreme}. The expected search cost grows quite slowly and is
approximately a constant plus $n/4$ for both problems over the range
of the figure.  Furthermore, the optimal number of steps, i.e., the
best value for $J$ in the algorithm also grows slowly. For the minimum
nogoods problem, $J$ ranges from 2 to 4 over this range, while the
maximum nogoods problem has $J$ ranging from about 5 to 15, and
appears to grow as $O(\sqrt{n})$. In both problems, the best number of
steps is considerably less than $L=n/2$.  The slow growth in search
cost for the maximum nogoods problem is particularly impressive since
an unstructured quantum search requires of order $\sqrt{n \choose L}$
steps, which for $n=100$ is about $3 \cdot 10^{14}$.  As a final
observation, if instead of \eq{phase}, we just invert the phase of
nogoods, the search cost is somewhat larger for the minimum nogoods
problem, and somewhat lower for the maximum nogoods problem.  Hence a
variety of phase adjustment policies have good performance for these
extreme problems.

\figdef{x:extreme}{ \epsfig{file=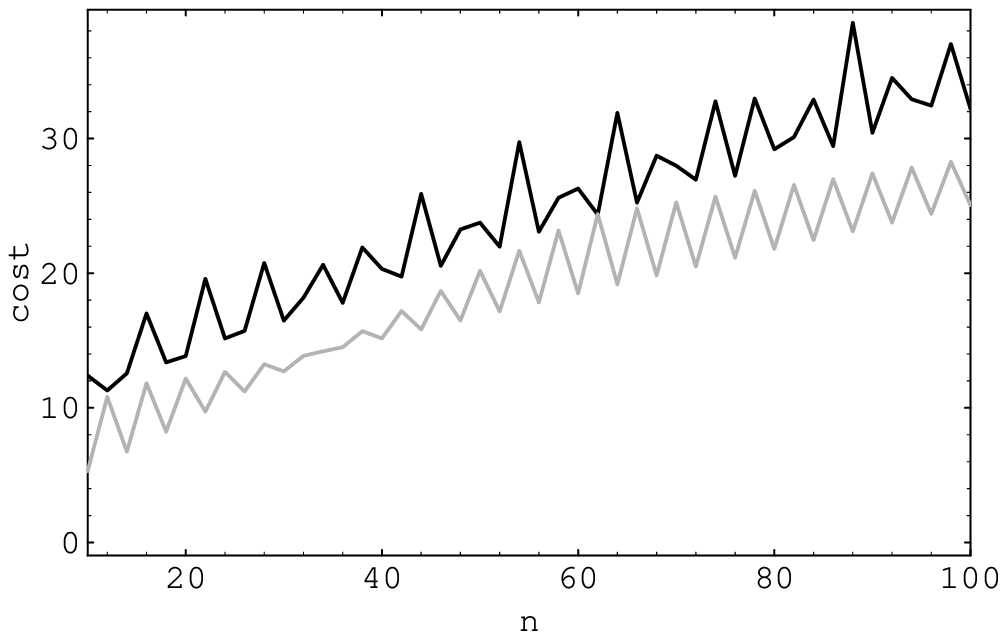} }{Expected
search cost for extreme problems for even values of $n$ with
$L=n/2$. Black and gray curves are for problems with the maximum and
minimum number of nogoods, respectively.}

\figdef{x:amplitude}{ \epsfig{file=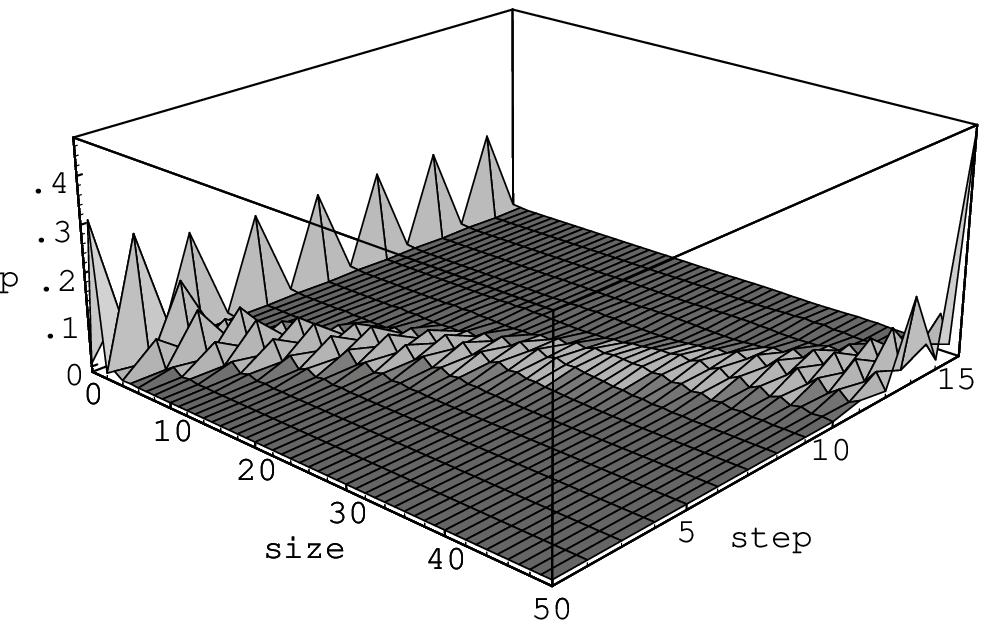} }{Evolution of
the probability in goods of different sizes for the maximum
nogoods problem with $n=100$ and $L=50$. Probabilities greater than
0.05 are shown with a lighter shade. After 15 steps,
the probability to have a solution, i.e., a good of size 50, is 0.47.}

Further insight into the behavior of this algorithm is given by
\fig{x:amplitude} which shows how the probability to have a good of
different sizes varies with each step of \eq{map}. Specifically, for
each step $j$ and each set size $k$, the figure shows the value of
$\sum_s |\psi^{(j)}_s|^2$ where the sum is over all good sets $s$ of
size $k$. Since the size of goods is at
most 50, the plot does not include larger sets. The initial condition
(not shown) has probability 1 in the set of size 0.
 The algorithm maintains a concentration of amplitude in
goods, and moves rapidly up the lattice to give a relatively large
amplitude in the solution after 15 steps, considerably fewer than $L=50$.

\subsection{Intermediate Cases: Hard Problems}

The difficult search problems, on average, have an intermediate number
of constraints: not so few that most complete states are solutions,
nor so many that any incorrect search choices can be quickly
pruned.

\figdef{x:example}{ \epsfig{file=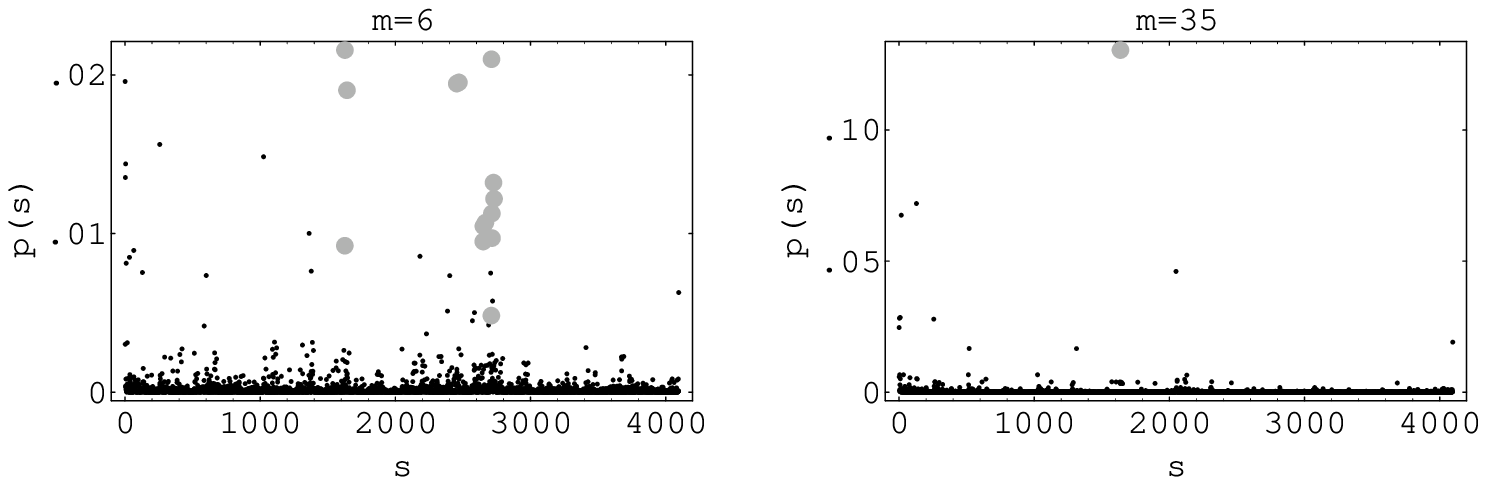} }{Probability
in each set for two CSPs with $n=12$ and $L=6$ after $J=5$
steps. On the left, the problem has $m=6$, with 14 solutions and
$P_{\rm soln}=0.19$, on the right, $m=35$, with a single solution and
$P_{\rm soln}=0.13$. The large gray points are the solutions.}

Two examples of how the algorithm concentrates amplitude into solutions
are shown in \fig{x:example}. For each problem, the figure shows the values
of $|\psi^{(J)}_s|^2$ for each of the $2^n$ sets in the lattice. The solutions
are drawn as gray points to distinguish them from the remaining sets.
For the plots, the sets are
ordered according to the integer whose binary representation corresponds
to including the items in the set.
For both of these problems, the lowest expected search cost is with $J=5$,
providing another illustration of the algorithm moving 
significant amplitude to the solution level of the lattice
in fewer than $L$ steps.

In these problems, the constraints are specified by nogood assignments of
size 2, corresponding to CSPs, such as graph coloring, where each contraint 
involves two variables. The value of $m$ denotes the number
of such nogoods in the problem. These two examples illustrate the typical
behavior for a problem with relatively few constraints, and many solutions,
and a problem with many constraints and only one solution.
There are
${12 \choose 6 }=924$ sets at the solution level, so a random selection
would give a probability of about 0.001 to each set, much less than
given to solutions by this algorithm. Thus the various contributions to
nonsolutions tend to cancel out among the many paths through the
lattice. The figure also illustrates the variation in $|\psi^{(J)}_s|^2$
among the sets showing that, unlike the extreme problems of the previous
section, the amplitudes do not depend only on the size of the set and
overlap with solutions. Rather the details of which constraints apply to
each set give rise to the variation in values seen here. This variation
precludes a simple theoretical analysis of the algorithm.

\subsubsection{Phase Transition}

For a more general indication of how this algorithm uses the structure
of search problems, we consider its average behavior for ensembles of
problems with different degrees of constraint. One such ensemble
consists of randomly generated instances of CSPs where the constraints
specify $m$ nogoods of size 2 and $L=n/2$. Increasing $m$ changes the
ensemble from weakly constrained to highly constrained problems, thus
showing how the performance depends on the tightness of the
constraints. Since the quantum algorithm can find solutions but never
definitely prove that no solution exists, we examine only problems
with at least one solution.

Specifically, for given values of $n$ and $m$, problem instances are
generated as follows. First, we randomly select a complete assignment
to be a solution for the problem. Then, from among the assignments of
size 2 that are not subsets of this prespecified solution, we pick $m$
distinct sets to be the nogoods directly determined from the problem's
constraints. All these problems also have necessary nogoods to
constrain each variable to have a unique solution. This generation
procedure guarantees the problem has at least one solution since it
never selects any subset of the prespecified solution to be a nogood.

Among the $n \choose 2$ sets of size 2, $4 {L \choose 2}$ are assignments
and the remainder are necessary nogoods. Of these assignments, $L \choose 2$
are subsets of the prespecified solution. The remaining sets are available
to be selected as the $m$ nogoods from the constraints. Thus to span the
range from unconstrained to fully constrained problems, we can select $m$
to range from 0 to 
\begin{equation}\eqlabel{mmax}
m_{\rm max} = 3 {L \choose 2}
\end{equation}

\figdef{x:transition}{
\epsfig{file=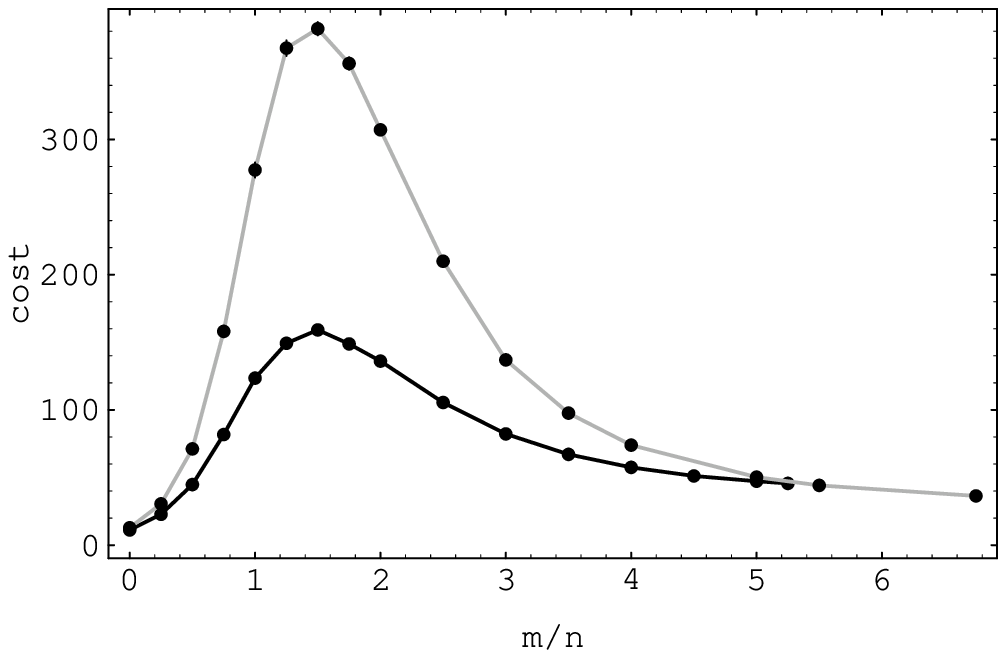,width=3.5in}
}{Average search cost $C$ as a function of $\alpha = m/n$ for $n=16$ and 20
(black and gray curves, respectively). Each point is the average of
1000 problem instances, and includes error bars indicating the
standard error of the mean, which are smaller than the
size of the plotted points.}

The average behavior of the algorithm for these problem ensembles as
the number of constraints are changed is shown in \fig{x:transition}.
This search algorithm exhibits the phase transition behavior described
above as occurring for many classical searches~\cite{cheeseman92}.
Thus the algorithm is using interference of paths to exploit problem
structure in the same manner as sophisticated classical search methods
are observed to do.

Because the location of the transition is at a value of $m$ that grows
linearly with $n$, the figure shows the search cost as a function of
$\alpha \equiv m/n$.  More precisely, a mean-field theory of this
behavior predicts the the transition point, and the peak in the search
cost, occur at $\alpha_{\rm crit} = -\ln(2)/(2 \ln(3/4)) = 1.2$ when
$n$ is large~\cite{williams92}. For the quantum search, the figure
shows the search cost peak is close to this value even for relatively
small values of $n$.

A significant observation from \fig{x:transition} is that the quantum
algorithm's search cost decreases after the transition. By contrast,
since the expected number of solutions continues to decrease as
problems become more constrained, unstructured search methods do not
show this decrease in cost. Thus the use of problem structure is
relatively more beneficial for problems with many constraints.

As with other examples of phase transitions in search, the variance in
search cost among different problems in the ensemble is relatively
large near the transition point. Thus an interesting open question,
for both classical and quantum search methods, is whether there are
simple ways to identify those problems likely to be much harder or
easier than the average. If these cases correspond to particular types
of structure~\cite{vlasie96}, it may be possible to develop
specialized variants of the search methods particularly well suited to
those cases.

\subsubsection{Scaling}

\figdef{x:scaling}{ \epsfig{file=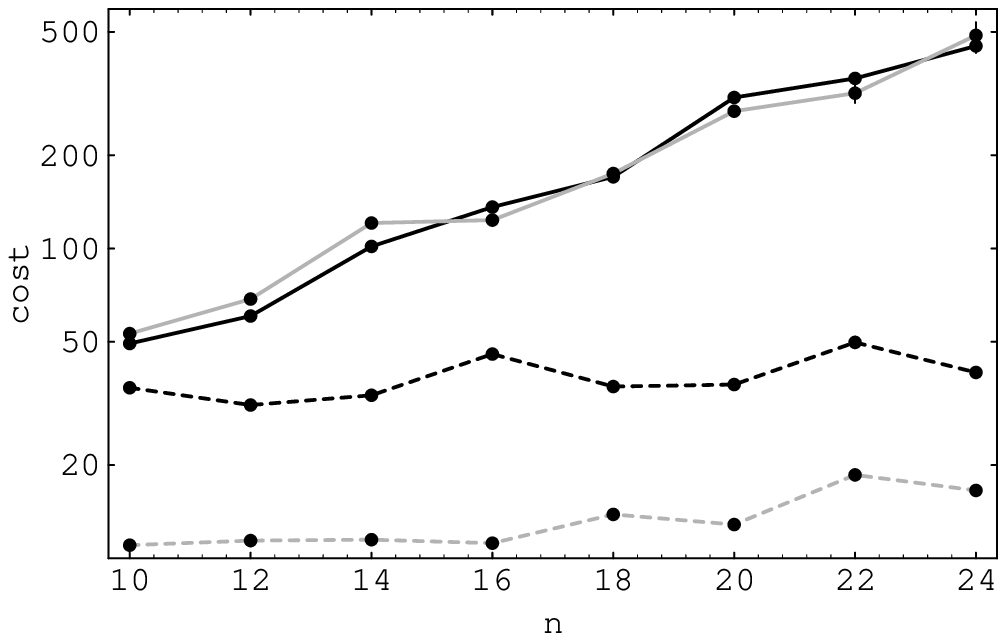} }{ Log plot of 
search cost $C$ vs.~$n$.  The solid curves are for
$\alpha=1$ (gray) and $\alpha=2$ (black).  Each point is
the average of 1000 CSPs, except for $n=22$ with 100 and $n=24$ with
50.  The points include error bars indicating the standard error of
estimates of the mean, which in most cases are smaller than the size
of the plotted points.  The dashed curves are for
 the CSPs with $m=0$ (gray) and with $m=m_{\rm max}$ of \eq{mmax} (black).
}

\fig{x:transition} shows the performance for two problem sizes.  An
important question is how rapidly the search cost grows with
increasing problem size. An appropriate choice of the scaling is
necessary for a study of average behavior so as to include a
significant number of hard instances.  In this respect, a useful
scaling regime is when the number of nogoods specified by the
constraints grows linearly with the size of the problem $n$. This
corresponds to graph coloring where the number of edges is
proportional to the number of nodes, and satisfiability where the
number of clauses in the propositional formula is proportional to the
number of variables, which have a high concentration of hard search
cases~\cite{cheeseman92,kirkpatrick94}.

The scaling of the search cost $C$ is shown in \fig{x:scaling}.
Although the problem sizes feasible for classical simulation may be
too small to see the asymptotic growth rate clearly, the search cost
appears to grow slowly but still exponentially, on average, for
$\alpha=1$ and 2.  Thus, the expected cost grows by about a factor of
10 while the full search space grows in size by a factor of $2^{14}$
over the range of the figure. These values of $\alpha$ correspond to
locations just below and just above the peak in the search cost show
in \fig{x:transition}.

The dashed curves in \fig{x:scaling} show the behavior for simpler
CSPs. In particular, the CSP with the fewest nogoods, i.e., $m=0$, has
a cost just slightly higher than the minimum nogoods problem discussed
in the previous section. Although $m=0$, the CSP still has the
necessary nogoods and so has more nogoods in the lattice than the
minimum nogoods problem. At the other extreme is the soluble CSP with
$m=m_{\rm max}$ of \eq{mmax}. Its search cost is about three times
larger than that of the maximum nogoods problem of the previous
section but shows the same slow growth in search cost. This CSP
differs from the maximum nogoods problem only in having no nogoods of
size 1, e.g., the sets $\{i\}$ for $i=L+1,\ldots,n$ are nogoods for
the maximum nogoods problem but not for the CSP with $m=m_{\rm max}$
when the prespecified solution is $\{1,\ldots,L\}$.

Overall, we conclude that this algorithm is very effective in
concentrating amplitude toward solutions, but it is unclear whether
that is enough to give polynomial rather than exponential decrease of
$P_{\rm soln}$, on average, for the hard problems near the transition.

\subsubsection{Comparison with Previous Algorithms}

\figdef{x:ratio scaling}{ \epsfig{file=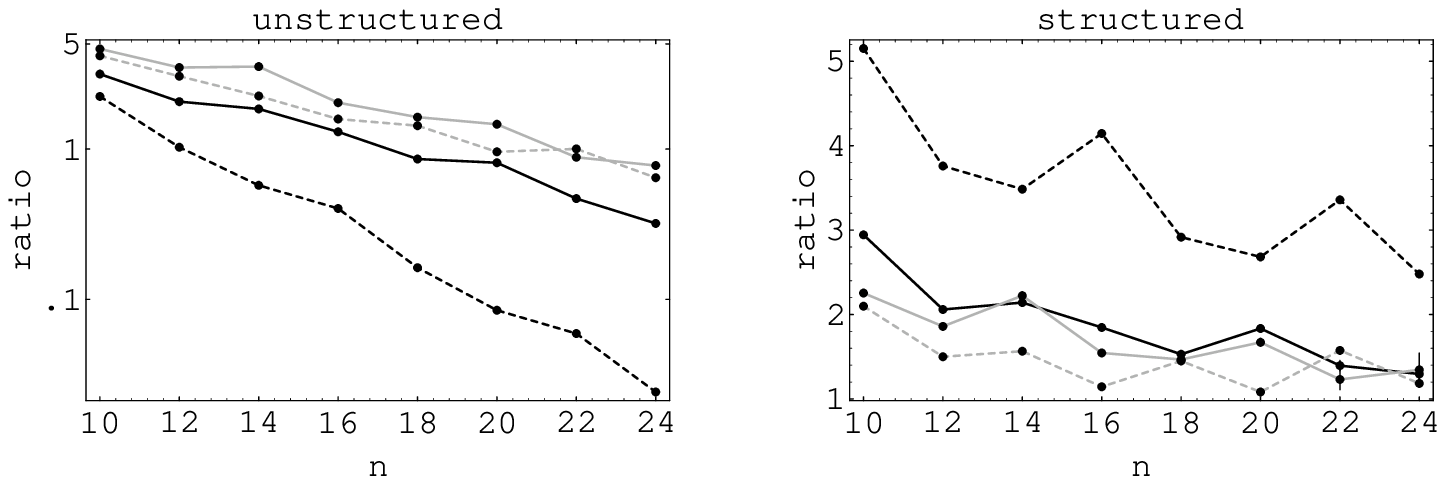} }{ Ratio of
$C$ for the new algorithm to the search cost scaling of previous ones.
The curves are for the same problems as used in \fig{x:scaling}.  The
left plot compares with the unstructured search method on a log scale,
and the right plot with the previous structured search on a linear
scale.}

Having introduced a new quantum algorithm, an important question is
how it compares with previous methods and, in particular, the types of
problems for which it is most appropriate. To address this question,
\fig{x:ratio scaling} shows how the search cost of this algorithm
compares with the scaling of the previous unstructured and structured
search methods.

Specifically, to compare with the unstructured method, the number of
solutions $S$ was found for each problem instance examined in
\fig{x:scaling}. The first plot of \fig{x:ratio scaling} shows the
ratio $C/\sqrt{N_L/S}$ averaged over the problem instances, where $N_L$
is the number of sets of size $L$.  The value $\sqrt{N_L/S}$
characterizes the scaling behavior of the unstructured search among
sets at the solution level of the lattice, although the actual search
cost may differ from this by a constant factor~\cite{boyer96}.  This
ratio generally decreases with $n$, especially for highly
constrained problems. If this trend continues for larger $n$, it would
mean the structured algorithm is able to improve search performance by
utilizing the problem structure, but definite conclusions cannot be
made from these small problem sizes.

The second plot of \fig{x:ratio scaling} compares the new algorithm to
the previous structured search algorithm. This previous algorithm is
similar in form to the new one, but uses a matrix $U$ that maps
entirely from one level of the lattice to the next at each step, and
whose elements have no simple closed form expression. Thus it requires
$J=L$ steps to have any amplitude in solution sets. For comparison,
this algorithm was run on the same problems as used in \fig{x:scaling}
and the expected search cost $L/P_{\rm soln}$ was determined for each
problem. The plot shows the ratio of the average search costs of the
new and old algorithms. We see that the new algorithm has slightly
higher cost, with the relative difference decreasing with $n$. The
difference between the two methods increases as problems become more
constrained, but this can be partly offset for highly constrained
problems by changing the phase adjustment of \eq{phase} to just invert
nogoods. As a further observation from the data used to generate the
plot, the two algorithms have comparable variance in search cost
within these problem ensembles. At any rate, for a small decrease in
performance, the new algorithm provides an analytically simpler search
method with the same qualitative behaviors as the previous structured
one.  This analytic simplicity may facilitate a theoretical analysis
of the new algorithm.

\section{ Discussion}

The algorithm presented here shows how the underlying lattice
structure of search problems can be used as the basis of a structured
quantum search algorithm. The algorithm is particularly effective for
relatively highly constrained problems. There remain a number of ways
in which the algorithm might be improved. First, the initial
motivation for the matrix $U$ was to maximally connect sets to their
immediate supersets. In fact, we found that the mapping allowed the
algorithm to work best with fewer steps than would be expected from
moving up one level at a time in the lattice. It may be possible to
design other mappings that do this even more effectively by somewhat
reducing the mapping to immediate supersets and increasing the
connections to larger supersets. 

Another issue is the structure of the types of mappings possible by
adjusting the phases of the diagonal matrix $D$. As we saw, this gives
matrix elements that depend only on the combination $|r|+|s|-2|r \cap
s|$ when the diagonal elements of $D$ depend only on the size of the
corresponding sets. Thus, for instance, there is no way to distinguish
mapping from a set to an immediate subset, i.e., moving away from the
solution level, from mapping to an immediate superset. For CSPs,
another limitation of this mapping is its inability to distinguish
necessary nogoods from other sets. Because these nogoods do not depend
on particular problem instances, it may be useful to have the matrix
$U$ emphasize moving amplitude not just to all immediate supersets
equally but rather to focus on those sets that are assignments. Such a
modification is likely to be most beneficial for problems with
relatively few constraints where the necessary nogoods are a high
proportional of all the nogoods in the problem. Thus considering a
wider possible range of mappings may allow a more focused search.

There are also a variety of phase adjustment policies. The one studied
here is quite effective, but other choices can enhance the performance
of the mapping.  Furthermore, since we focus on typical or average
behavior, other choices that do not improve the average but result in
smaller variance would also be useful in improving the predictablility
of the algorithm's performance.

As a possible extension to this algorithm, it would be interesting to
see whether the nonsolution sets with relatively high probability
could be useful also, e.g., as starting points for a local repair type
of search~\cite{minton92}. If so, the benefits of this algorithm would
be greater than indicated by its direct ability to produce solution
sets. This may also suggest similar algorithms for the related
optimization problems where the task is to find the best solution
according to some metric, not just one consistent with the problem
constraints.

Beyond the specific algorithm presented in this paper, the lattice
structure provides a general framework for applying quantum computers
to search problems. This is due to the many opportunities for using
interference among the paths through the lattice to each set at the
solution level.  Specifically, this search framework could be used to
incorporate additional knowledge about the particular problem
structure or other search heuristics. This is readily included as a
modification to the choice of phases which can be made independently
for each state.  For example, the choice of phase adjustment policy
could be based on the number of constraints in a given problem, i.e.,
using \eq{phase}, unless the problem is highly constrained.  In this
latter case, we could just invert nogoods only, resulting in somewhat
better performance.

Changes to the mapping that mixes amplitude among sets are more
complicated due to the requirement to maintain unitarity (as well as
computational simplicity). Nevertheless, when a heuristic works well
classically for a particular class of problems, it suggests a
corresponding unitary mapping that is as close as possible to the
classical procedure the heuristic uses to move from one state to the
next during search.  This method for constructing quantum search
mappings was the underlying motivation for the specific map used in
the structured search presented here.  For example, it would be
interesting to examine maps motivated by classical repair and
abstraction search methods.

Another way to incorporate heuristics is by changing the initial
condition. In the method reported here, initially all amplitude is in
the empty set.  Other possibilities include starting with amplitude in
the consistent sets at the level of the lattice corresponding to the
nogoods directly determined by the constraints~\cite{hogg95d} or
starting with an equal superposition in all sets of the
lattice~\cite{grover96}.

There remain a number of important questions. First, how are the
results degraded by errors and decoherence, the major difficulties for
the construction of quantum
computers~\cite{landauer94a,unruh94,haroche96,monroe96a}? While there
has been recent progress in
implementation~\cite{barenco95,cirac95,cory96,gershenfeld96,sleator95},
quantum approaches to error control~\cite{berthiaume94,shor95} and
studies of decoherence in the context of factoring~\cite{chuang95} it
remains to be seen how these problems affect the framework presented
here.

Second, it would be useful to have a theory for asymptotic behavior of
this algorithm for large $n$, even if only approximately in the spirit
of mean-field theories of physics. This would give a better indication
of the scaling behavior than the classical simulations, necessarily
limited to small cases, and may also suggest better phase
choices. Considering these questions may suggest simple modifications
to the quantum map to improve its robustness and scaling.  There thus
remain many options to explore for using the deep structure of
combinatorial search problems as the basis for general quantum search
methods.

\section*{Acknowledgements}

I have benefited from discussions with S. Ganguli, J. Gilbert and C. Williams.

\end{document}